\documentclass{article}

\usepackage{graphicx}
\usepackage{multirow}
\usepackage{subfig}
\usepackage{multicol}
\usepackage{amsmath}
\usepackage{pifont}

\usepackage{algorithm}
\usepackage[noend]{algpseudocode}

\newcommand{\cmark}{\ding{51}}
\newcommand{\xmark}{\ding{55}}

\begin{document}

\providecommand{\keywords}[1]
{
  \small	
  \textbf{\textit{Keywords---}} #1
}

\title{Understanding Money Trails of Suspicious Activities in a cryptocurrency-based Blockchain}

\author{Banwari Lal,
        Rachit Agarwal, 
        Sandeep K. Shukla 
    \\
    CSE, IIT Kanpur, India,\\
    Email: \{banwari, rachitag, sandeeps\}@iitk.ac.in
}

\maketitle
\begin{abstract} 

The decentralization, redundancy, and pseudo-anonymity features have made permission-less public blockchain platforms attractive for adoption as technology platforms for cryptocurrencies. However, such adoption has enabled cybercriminals to exploit vulnerabilities in blockchain platforms and target the users through social engineering to carry out malicious activities. Most of the state-of-the-art techniques for detecting malicious actors depend on the transactional behavior of individual wallet addresses but do not analyze the money trails. We propose a heuristics-based approach that adds new features associated with money trails to analyze and find suspicious activities in cryptocurrency blockchains. Here, we focus only on the cyclic behavior and identify hidden patterns present in the temporal transactions graphs in a blockchain. We demonstrate our methods on the transaction data of the Ethereum blockchain. We find that malicious activities (such as Gambling, Phishing, and Money Laundering) have different cyclic patterns in Ethereum. We also identify two suspicious temporal cyclic path-based transfers in Ethereum. Our techniques may apply to other cryptocurrency blockchains with appropriate modifications adapted to the nature of the crypto-currency under investigation.  

\end{abstract}


\section{Introduction}\label{sec:intro}

Blockchain technology works on a peer-to-peer (P2P) overlay network over the Internet, employs cryptographic algorithms to secure the record of transactions, and ensures transactions are valid and authorized. Permission-less blockchain-based applications are distributed, redundant, and provide their users with a sense of anonymity. While this benefits privacy, it also allows malicious actors to hide their true identities and perform illegal transactions. The anonymity in most common blockchain-based cryptocurrency platforms is not full proof, and transaction data correlation has effectively apprehended some malicious actors in the past. However, such correlation requires processing massive historic transaction datasets, creating graph databases, and implementing search mechanisms specific to the data. Therefore, Law Enforcement Agencies (LEAs) are not always capable of identifying malicious actors. Knowing this, malicious actors carry out illicit activities.  Such misuse has, to date, resulted in a considerable loss in cryptocurrencies such as Bitcoin and Ethereum~\cite{bitcoinAttackSurvey, scAttackSurvey}. Cryptocurrencies are among the most preferred forms of exchange for illicit transactions pertaining to dark web markets, ransomware attacks, Phishing, Gambling, and laundering funds by cyber-criminals. In~\cite{chainAna}, the authors show that Phishing generates more than 50\% of the cyber-crime revenue. Such cyber-attacks and activities exploit the vulnerabilities present in the blockchain infrastructure or target its users through social engineering. Thus, a critical issue that needs attention is to secure users from cyber-attacks yet maintain their privacy.

Different state-of-the-art approaches such as~\cite{malDetect,agarwal2020detecting} exist that detect suspicious accounts in cryptocurrency-based blockchains. However, these techniques either consider only static features extracted from an aggregated graph of user interactions in the blockchain~\cite{malDetect} (thereby neglecting behavioral changes over time) or use temporal features to identify behavioral changes and ignore different types of malicious activities~\cite{agarwal2020detecting} (i.e., consider all the malicious activities under one class). In~\cite{agarwal2021detecting}, the authors show that different malicious activities present in the Ethereum blockchain can be clustered in 4 clusters. These 4 clusters contain accounts related to malicious activities such as \textit{Phishing}, \textit{Gambling}, \textit{Money laundering}, and \textit{others}, respectively. However, the authors consider only a few properties that were extracted after considering the ``local neighborhood'' (a local neighborhood comprises of the accounts to which a particular account has transacted with) to identify clusters. Further, they do  not comment on why they obtain such results. In our view, the results in~\cite{agarwal2021detecting} depend on how Phishing, Gambling, Money laundering, and other malicious activities are performed and how the money flow happens between the accounts involved in these activities. 

Thus, the question we ask is, \textit{\textbf{Q: Does analyzing the money trails help distinguish the different malicious activities}}? To answer the question, in~\cite{motifs,motifs1, motifs2}, the authors consider Motifs (basic building blocks that repeat themselves in a graph) and two-hop cycles to understand motif applications in the temporal transaction graph of a blockchain. However, they do not consider the temporal cycles of larger lengths. In another work, in~\cite{financialfraud}, the authors show  that detecting money trails and understanding temporal user interactions help identify money laundering-based frauds that exist in finance-based institutions. Considering such aspects and the advancements in the state-of-the-art approaches, we aim to understand the differences in the various classes of malicious activities (those mentioned above) regarding how money flow happens in cryptocurrency-based blockchains and how we can use these paths to identify malicious activities. Thus, in this research, our goal is to understand the behavior of different malicious activities (e.g., Phishing, Gambling, and Money Laundering) in cryptocurrency-based blockchains such as Ethereum by understanding the money trail (a path followed by a cryptocurrency) Note that our approach is generic and can extend to any permission-less blockchain, with platform-specific adaptations.
 
We propose a money trail-based approach and analyze accounts involved in different malicious activities. Using the temporal cycles, we distinguish malicious activities such as Phishing, Gambling, and Money laundering. We use money loss (amount of cryptocurrency lost along a cyclic path) and the cyclic path's structure to understand the behavior of malicious accounts in the blockchain. We find that it is possible to detect suspicious behavior using money trails, and all the accounts involved in Phishing-based activities do not behave similarly. Further, some of the cyclic transfers result in high money loss in a short time. In short, our contributions are:

\begin{enumerate}
    \item We propose an \textbf{approach} based on money trails using temporal cycles:
    \begin{enumerate}
        \item to characterize the behavior of malicious activities such as Phishing, Gambling, and Money laundering in a cryptocurrency based blockchain,
        \item to detect suspicious temporal cycle-based money transfers using money loss along the cyclic path.
    \end{enumerate}
    \item As a \textbf{result}, we find that most malicious activities in Ethereum have similarity and can be clustered, while Phishing based accounts clusters themselves into more than one cluster. Also, using money trails, we find two suspicious temporal cyclic path-based cryptocurrency transfers in the blockchain.
    \item Current state-of-the-art approaches either use a machine learning or graph embedding-based approach to detect and analyze suspicious behavior in the blockchain. Our approach presents a \textbf{new dimension} to the existing approaches to study malicious activities using money trails.
    
\end{enumerate}

The structure of this article is organized as follows. Section~\ref{sec:back} and Section~\ref{sec:rw} describe relevant background material necessary for addressing the question {\em {\bf Q}} and illustrate the state-of-the-art approaches in search of answering this question, respectively. Section~\ref{sec:method} details the graph construction, our assumptions, and the methodology used in tracking the money trails. Section~\ref{sec:eval} presents the details of the data we use to validate our approach, how we collect the data, statistics about the data, and our experimental results. Section~\ref{sec:conclusion} finally conclude our paper and discuss future directions.

\section{Background}\label{sec:back}

Blockchain technology presents a decentralized (no central authority) data repository  of a digital ledger of transactions. The key difference between a typical database system and a blockchain ledger is how the data is structured. A ledger records transactions in a way that makes it difficult or impossible to change the information present in a transaction. Blockchain technology ensures that  the ledger is almost impossible to tamper with, and the integrity of the entries in the ledger  is cryptographically ensured~\cite{blockChain}. In a ledger, transactions are grouped to form ``blocks'' where each block contains many transactions. These blocks are chained together to construct a ledger. Each block in the ledger contains the details of the transactions included in the block, its own hash value, and most importantly, the previous block's hash value and a timestamp.  The decision on which block to include in the ledger among many competing blocks created by multiple participants is taken through a ``consensus algorithm'' (it verifies transactions, makes sure that the majority of members agree on the block to be added to the blockchain, and decides the current state of the ledger). In a blockchain, each new block  (to be included in the ledger) is duplicated and distributed across the entire ``blockchain network'' (the ledger is duplicated across a network of computers instead of storing at a central location). Each member of a blockchain network has an identical copy of the ledger. 

There are primarily two types of blockchains:
\begin{enumerate}
    \item A \textbf{Private blockchain} is a restricted blockchain where only a closed group of users are allowed to use the blockchain. If the right to transact or create blocks is restricted to permitted participants, then such a  blockchain is also called a ``permission-ed blockchain.'' Such blockchains are used within a consortium of organizations or enterprises that control accessibility and authorization. Some example applications of a private blockchain are blockchains supporting voting systems and digital identity.
    
    \item A \textbf{Public blockchain} is a  distributed ledger system which can be accessed by everyone. Here, users are allowed to transact and add blocks without any permission. Such blockchains are also called the permission-less blockchain. Here, there is no central authority to control the access and authorization of participants. A user can join or leave the permission-less blockchain whenever they want. A user who is part of a permission-less  blockchain can verify transactions, access current and past records, or do ``mining'' (a process of transaction verification and update of the records on the blockchain ledger).

\end{enumerate}
Although there are two more types of blockchains: consortium-based and hybrid blockchains, these blockchains are derived using the features of public and private blockchains. We do not describe them here.

An activity that is undertaken to cause harm to someone or carry out activities banned by law in certain jurisdictions is called a \textbf{malicious activity}. Some of the most prominent malicious activities are Phishing, Gambling, and Money Laundering. Although there are other activities such as Scamming and Heist, we only focus on Phishing, Gambling, and Money Laundering in this paper. The reason for focusing on these three is that the results in~\cite{agarwal2021detecting} show that most of the accounts involved in many malicious activities in Ethereum show statistically significant behavioral similarities to behaviors observed in accounts of these three activities. Note that these results are based on the available ground truth about the marked malicious account. Accounts related to many malicious activities (such as darknet marketplaces) are not marked and thus are out of the scope. 

Phishing is an activity where a cyber-criminal tries to steal digital cryptocurrency or user's credentials by using social engineering methods (for example, creating a fake website for wallets that look similar to the original website or send a fake email)~\cite{phishScamdetection}. In Phishing activity, attackers do not exploit the vulnerabilities present in the system but trick the human mind's inattention. Here, attackers aim to get sensitive information, install the malware in the system, or steal digital currency. One of the widely known Phishing attacks on blockchains was the ``Bee Token ICO Scam'' attack~\cite{beeToken}. In the Bee Token ICO scam, the attackers got hold of emails associated with the accounts related to the Bee token and sent out emails to transfer Eth (a cryptocurrency used by Ethereum).

Gambling~\cite{gambling} is a process of taking part in an activity in which a person risks his money or a valuable object on the presumption of a specific outcome of an  event that induces uncertainty of  monetary loss or gain. Some Gambling activities are a lottery, video lottery games, card games, and casino games. Due to anonymity and transparency in the transactions, cryptocurrency blockchains are widely used in Gambling-related activities. Some of the main advantages of using cryptocurrency for Gambling are:
\begin{itemize}
    \item Chances of fraud are less due to the immutability of the blockchain.
    \item Anonymity allows users to participate in Gambling from   places where Gambling is legally prohibited.
    \item The transaction fee is small.
\end{itemize}

Some online Gambling platforms such as Las Atlantis Casino and Wild Casino for casino games also accept cryptocurrencies.

On the other hand, Money Laundering~\cite{bitCoinAML} is the process of obfuscating the true source of an income by moving the money into various hands, through mixers, and eventually to get it returned. At the end of such obfuscating transfer cycle, it seems that the  money is transferred from a legitimate source. Some of the money laundering-based criminal activities are drug trafficking, and terrorist funding, etc. Although it is easier to do illicit activities in cryptocurrency due to user anonymity in blockchain, the overall impact of cryptocurrencies on money laundering is less compared to cash transactions. As of 2019, ``only 0.5\% of all Bitcoin transactions''~\cite{bitcoinEliptic} involved trading Bitcoin on the dark web. In money laundering, criminals hide the true origin of illicit funds using various methods (such as participating in Initial Coin Offerings (ICO) and converting one type of coin into another). In~\cite{bitCoinAML}, the authors state three main stages of money laundering in cryptocurrency. These stages are:
\begin{enumerate}
    \item \textit{Placement:} Cash or other types of crypto (altcoin) can be used to purchase cryptocurrency from exchanges. A legal transaction requires identity verification, identification of  the fund source, and following Anti Money Laundering Practices (AML). Exchanges that cannot force AML practices with sub-par tools and fails to check specific identity, allow exploitation of vulnerabilities and thereby enable money laundering. Certain exchanges have been found to be involved in such practices. 
    \item \textit{Hiding:} Once the digital currency is in play, criminals use the anonymizing aspect of cryptocurrency blockchains  to hide the source of the money.
    \item \textit{Integration:} In this phase, the criminals declare the money as a result of a profitable venture or other cryptocurrency appreciation to legitimize the illicit money.
\end{enumerate}

\section{Related Work}\label{sec:rw}

Several state-of-the-art approaches detect malicious accounts associated with such illicit activities in cryptocurrency blockchains, especially in Bitcoin and Ethereum~\cite{bitcoinAnomaly,bitcoinAttackSurvey,frudSurvey,agarwal2020detecting,agarwal2021detecting,malDetect1}. These state-of-the-art approaches either detect malicious activities using machine learning (ML) algorithms or by analyzing graphs using various metrics. Here, we first discuss  state-of-the-art approaches using ML and then   the graph analysis-based approaches. We finally demonstrate the state-of-the-art approaches for finding temporal cycles in a temporal graph. 

ML-based approaches use the features extracted from the underlying social interaction network. These features include degree (both inDegree and outDegree), transaction fee, balance, and clustering coefficient. Some of these approaches use the aggregated blockchain graph~\cite{bitcoinAnomaly}, while some use temporal graphs~\cite{agarwal2020detecting}. In~\cite{bitcoinAnomaly}, the authors use two aggregated graphs to detect suspicious users and transactions in Bitcoin. One graph has users as nodes for detecting suspicious users, and the other has transactions as nodes for detecting suspicious transactions. The approach uses features based on degrees (inDegree, outDegree, uniqueInDegree, uniqueOutDegree, and clustering coefficient), average (in-transaction, out-transaction), and balance. The authors used three unsupervised ML techniques (\textit{a}) K-means clustering, (\textit{b}) Mahalanobis Distance-Based Method, and (\textit{c}) Unsupervised SVM to detect suspicious user/transaction. In~\cite{bitcoinAnomaly}, the authors detect two users and one transaction out of the known suspicious user/transactions in the bitcoin. However, in~\cite{bitcoinAnomaly}, the authors do not use temporal aspects of blockchain's underlying social interaction network.

In~\cite{agarwal2020detecting}, the authors propose an ML-based approach that uses graph-based temporal features (such as burst and attractiveness) inspired by past attacks on a blockchain. They show  that indegree and outdegree present in the social interaction network of blockchain transactions follow power-law. The authors achieve balanced accuracy $\geq 87.2\%$ using ExtraTreeClassifier towards detecting malicious accounts. Nonetheless, when using the unsupervised learning approach, they detect  814 new malicious accounts that have a high probability of being malicious over time. However, in~\cite{agarwal2020detecting}, the authors consider all the malicious activities under one class. Thus, their results are biased towards a malicious activity with the most number of tagged accounts~\cite{agarwal2021detecting}. In~\cite{agarwal2021detecting}, the authors show that the Neural Network model  is resistant to such bias and detects any adversarial data. Further, they also show  that most malicious activities can be clustered into four clusters. However, they do not specify the reason behind their results and why four clusters are obtained. 

There are a few state-of-the-art approaches that focus on specific malicious activities~\cite{ponziScheme,ponziScheme1}. In~\cite{ponziScheme}, the authors focus on the ``Ponzi scheme''~\cite{ponziCFI} to detect illicit behavior using supervised ML techniques such as Random Forest Classifier, Bayes Network Classifier, and RIPPER Classifier. They handle the class imbalance problem using the oversampling (minority class's data is replicated) and the cost-sensitive (misclassification of minority classes is penalized more) approaches. They find the combination of cost-sensitive approach and Random Forest classifier provides the best accuracy~\cite{ponziScheme}. In~\cite{ponziScheme1}, the authors use ML to identify the smart contracts  involved in the Ponzi scheme in Ethereum. Their approach uses features based on user account and those from op-codes of the smart contracts. However, being dedicated to a specific malicious activity, these approaches do not apply to other attacks due to differences between different malicious activities.

In~\cite{ethGraphAnalysis}, the authors perform a graph-based analysis of the transactions present in the Ethereum blockchain. They use three graphs (\textit{a}) Money Flow Graph (MFG): a directed graph for money transfer based transactions, (\textit{b}) Smart Contract Creation Graph (CCG): a directed graph where an edge (\textit{u,v}) represents the node \textit{u} creating contract node \textit{v}. Here, node type is either Smart Contract (SC) or Externally Owned Account (EOA), and (\textit{c}) Contract Invocation Graph (CIG): a directed graph where an edge (\textit{u,v}) indicates node \textit{u} invokes the contract node \textit{v}. After analyzing the 3 graphs using different metrics (such as degree distribution, clustering, node connectivity, strongly/weakly connected component), the authors find that only a few SCs are dominant. The authors also propose an approach to address two security issues (``attack forensics'' - for a given malicious smart contract, finding all accounts controlled by the attacker, and ``anomaly detection'' - detection of abnormal contract creation) on Ethereum blockchain using cross graph analysis. However, the authors do not comment on any specific malicious activity.

In~\cite{tEdge}, the authors propose a Temporal Weighted Multidigraph Embedding (T-EDGE) approach based on graph embedding to classify Phishing/non-Phishing accounts. The approach uses the temporal weighted multi-edged directed graph where nodes represent unique accounts in Ethereum and the edges are temporal and contain details related to the transactions. The role of the graph embedding is to detect implicit features of the accounts in the transaction network of the Ethereum blockchain. The authors use two baseline embedding methods for the classification of Phishing/non-Phishing accounts. As one baseline method, they use a random temporal walk on the dynamic transaction graph to find the comprehensive properties (structural relationship) between accounts. This type of random walk contains money flow information in the blockchain. As a second baseline method, the authors use the skip-gram~\cite{skipGram} model to update the process parameters in the Neural Network using the Stochastic Gradient Descent algorithm. Using T-EDGE, the authors show that time-dependent walks and edge information are essential in a time-dependent transaction network. Further, they show that their approach achieves accuracy $\geq82\%$. However, their approach is dedicated to a particular malicious activity, and no insights are provided on the applicability of the approach to other malicious activities.

In~\cite{motifs1}, the authors propose an approach based on motifs (a subgraph that repeats themselves in a network or group of networks) to find ``mixing services''~\cite{bitcoinMagzine} (services that allow users to mix their coins with those of other users to enhance the anonymity of the transactions) in the Bitcoin blockchain. One of the mixing services prevalent in the financial sector and cryptocurrency-based blockchains is Money laundering. In~\cite{motifs1}, the authors used two types of temporal directed transactions graphs (\textit{a}) homogeneous Address-Address Interaction Network (AAIN) and (\textit{b}) heterogeneous Transaction-Address Interaction Network (TAIN). The approach uses hybrid motifs composed of temporal homogeneous motifs in AAIN and attributed temporal heterogeneous (ATH: local subgraphs of heterogeneous information network) motifs in TAIN. In~\cite{motifs1}, the authors detect $\geq 6.4\%$ of total addresses as mixing addresses using hybrid motifs as a key feature. Their detection model is based on the ground-truth information, thus unknown complex mixing strategies that may exist might not be detected. In~\cite{financialfraud}, the authors propose an approach for detecting frauds using money trails in the temporal transactions graph. They integrate their approach with a real-time fraud detection system in a private bank. This approach was able to detect 2-4 new illicit activities every month since 2017. The approach in~\cite{financialfraud} also allowed banks to set new preferences (for example, the limited time between transactions or maximum percentage of money loss is allowed).

The state-of-the-art approaches related to general cycle detection~\cite{elementrycircuit,simpleCycles,2scent} are based on a depth-first search (DFS) algorithm~\cite{dfs}. In~\cite{elementrycircuit}, the authors propose an approach to find all the elementary cycles present in a graph using the DFS. However, the approach does not consider temporal aspects for finding elementary cycles. On the other hand, in~\cite{simpleCycles,2scent}, the authors provide methods to find all the elementary temporal cycles in a temporal graph. Nonetheless, these approaches do not apply other preferences (for example, the maximum percentage of money loss allowed). Such aspects inhibit us from using these approaches for our problem as, besides timestamps, transaction amount-based attributes are also important.

In summary, state-of-the-art approaches related to illicit activity detection in the blockchain using ML and graph embedding are well studied. However, most ML-based approaches are either biased towards a malicious activity with a large number of tagged  accounts/transactions or do not consider temporal aspects. Also, most ML-based approaches do not use hidden patterns produced (such as cycle-based transfers) by the transactions network of blockchain. Thus, in our view, an approach to detect money trails for analysis of different malicious activities needs to be  formalized. This paper is our attempt to do precisely that. 

\section{Methodology}\label{sec:method}

This section provides a detailed description of our methodology towards identifying time-respecting cycles to detect money flow.

\subsection{Cycle Detection}

\begin{figure}
    \centering
    \subfloat[Example Graph]{
        \includegraphics[width=0.45\textwidth]{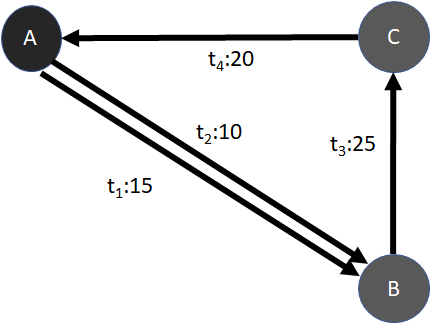}
        \label{fig:assumption1}
    }
    \subfloat[Graph after applying first modification]{
        \includegraphics[width=0.45\textwidth]{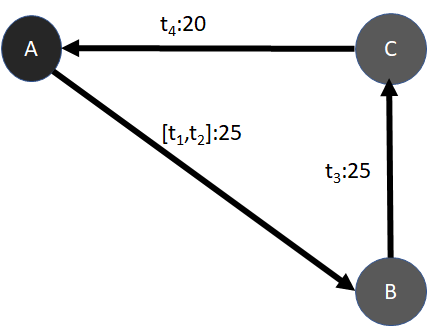}
        \label{fig:afterassumption1}
    }\\
    \subfloat[Example Graph considered for second modification]{
        \includegraphics[width=0.48\textwidth]{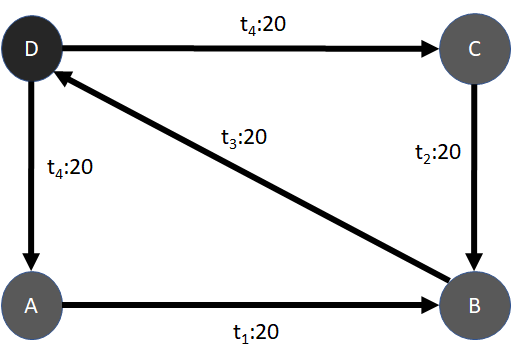}
        \label{fig:assumption2}
    }
    \subfloat[Example Graph considered for third modification]{
        \includegraphics[width=0.48\textwidth]{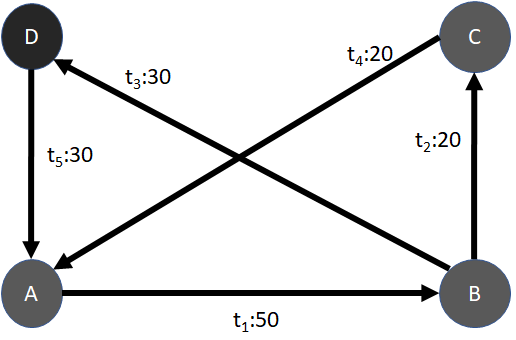}
        \label{fig:assumption3}
    }
    \\
    \subfloat[Example Graph towards all modifications]{
        \includegraphics[width=0.65\textwidth]{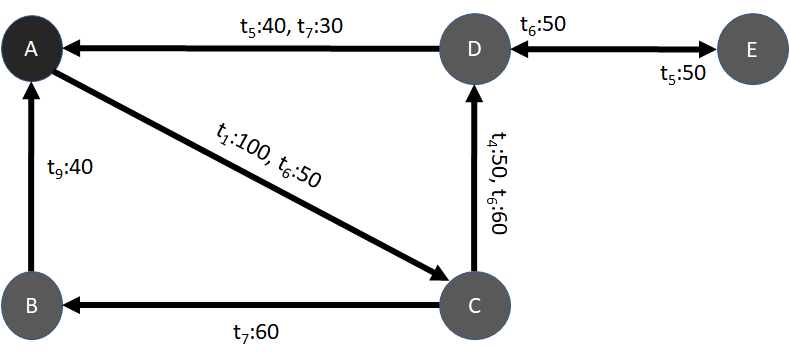}
        \label{fig:allSpecialCases}
    }
     \caption{Cases considered for modifications}
    \label{fig:specialCases}
\end{figure}

The standard cycle detection methods do not consider the temporal aspects. Thus, to find time-respecting cycle to detect money flow, there is a need to modify the cycle detection method. In a temporal cycle the order of the edges in a temporal path is restricted by the time in which they occur and can have only one starting node. However, in standard cycle detection methods an edge can be involved in more than one cycle. Thus, it is essential to identify all such cases in which two or more cycles contain common edge/edges. Figure~\ref{fig:specialCases} shows an example of such cases and depicts four different temporal graphs where an edge $A\xrightarrow{t_i:x}B$ in the graph represents that a user $A$ has transferred $x$ contains the other details such as amount transferred, gas price incurred. For now for simplicity, consider for $x$ to be amount of cryptocurrency transferred to user $B$ at time $t_i$. Our method is based on following special cases.

\begin{enumerate}
    \item Figure~\ref{fig:assumption1} shows two temporal cycles (\textit{a}) $A\xrightarrow{t_1:15}B\xrightarrow{t_3:25}C\xrightarrow{t_4:20}A$ and (\textit{b}) $A\xrightarrow{t_2:10}B\xrightarrow{t_3:25}C\xrightarrow{t_4:20}A$ in the graph where the edges $B\xrightarrow{t_3:25}C$ and $C\xrightarrow{t_4:20}A$ are common in both the temporal cycles. All the incoming edges from node A to node B have less timestamp than outgoing edges from B in the temporal cycle. So instead of considering two different edges from node A to node B, we merge both the edges from node A to node B into one edge, in this case, $A\xrightarrow{t_1,t_2:25}B\xrightarrow{t_3:25}C\xrightarrow{t_4:20}A$. Here, we add other information like amount of cryptocurrency transferred via both the edges. Figure~\ref{fig:afterassumption1} shows the graph after our modification.
    
    \item Figure~\ref{fig:assumption2}, although different, again shows two temporal cycles (\textit{a})
    $A\xrightarrow{t_1:20}B\xrightarrow{t_3:20}D\xrightarrow{t_4:20}A$ and (\textit{b}) $C\xrightarrow{t_2:20}B\xrightarrow{t_3:20}D\xrightarrow{t_4:20}C$ in the graph. Here, one temporal cycle starts from node A and the other from node C. But the timestamp of the edge from node A is less than the edge from node C. We give the edge from node with a lower timestamp a higher priority  because that edge (transaction) has happened earlier in the blockchain. If the timestamp of two edges is equal, then the edge that transfers higher cryptocurrency or pays higher gas price gets higher priority. Thus, in the example, the temporal cycle originating from node A gets priority.
    
    \item Further, Figure~\ref{fig:assumption3} shows two temporal cycles (\textit{a}) $A\xrightarrow{t_1:50}B\xrightarrow{t_2:20}C\xrightarrow{t_4:20}A$ and (\textit{b}) $A\xrightarrow{t_1:50}B\xrightarrow{t_3:30}D\xrightarrow{t_5:30}A$ in the graph. Edge $A\xrightarrow{t_1:50}B$ is common in both the cycles. If we consider up to timestamp $t_4$, money recovered at account A (incoming money in the temporal cyclic path with respect to outgoing money) is only 20 cryptocurrency. However, when considering $t_5$, the remaining money is recovered by account A. Thus, we consider both the temporal cycles. Note that this also depends on the data we have. 
\end{enumerate}

\begin{figure}
    \centering
    \includegraphics[width=0.8\textwidth]{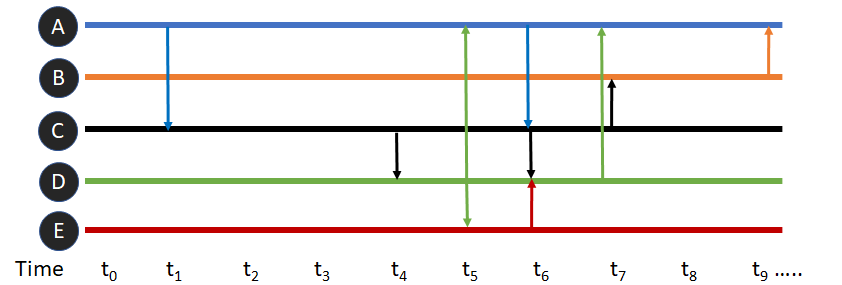}
    \caption{Contact Sequence of graph represented in Figure~\ref{fig:allSpecialCases}.}
    \label{fig:contactsq}
\end{figure}

\begin{table}
    \centering
    \begin{tabular}{|l|l|l|}
    \hline
     \textbf{\#} & \textbf{All Temporal Cycles} & \textbf{Temporal Cycles After Modifications}  \\
     \hline
        1 & $A\xrightarrow{t_1:100}C\xrightarrow{t_4:50}D\xrightarrow{t_5:40}A$ & \multirow{3}{*}{$A \xrightarrow{t_1:100} C\xrightarrow{t_4,t_6:110}D \xrightarrow{t_5,t_7:70}A$ }  \\
        \cline{0-1}
        
        2 & $A\xrightarrow{t_1:100}C\xrightarrow{t_4:50}D\xrightarrow{t_7:30}A$& \\
        \cline{0-1}
        3 & $A\xrightarrow{t_1:100}C\xrightarrow{t_6:60}D\xrightarrow{t_7:30}A$ & \\
        \hline
        4 & $A\xrightarrow{t_1:100}C\xrightarrow{t_7:60}B\xrightarrow{t_9:40}A$ &\multirow{2}{*}{$A\xrightarrow{t_6:50}C\xrightarrow{t_7:60}B\xrightarrow{t_9:40}A$ }  \\
        \cline{0-1}
        5 & $A\xrightarrow{t_6:50}C\xrightarrow{t_7:60}B\xrightarrow{t_9:40}A$ & \\
        \hline
        6 & $D\xrightarrow{t_5:50}E\xrightarrow{t_6:60}D$ & $D\xrightarrow{t_5:50}E\xrightarrow{t_6:60}D$ \\
        \hline
        
    \end{tabular}
    \caption{List of Cycles before and after applying our modifications}
    \label{tab:cyclesList}
\end{table}

Figure~\ref{fig:contactsq} shows the contact sequence of the graph present in Figure~\ref{fig:allSpecialCases}. Table~\ref{tab:cyclesList} shows all the 6 temporal cycles present in this graph that meet our criteria and the remaining 3 cycles after the application of our modifications mentioned above. The 3 temporal cycles start at 3 distinct timestamps (\textit{a}) $A\xrightarrow{t_1:100}C$, (\textit{b}) $A\xrightarrow{t_6:50}C$, and (\textit{c}) $D\xrightarrow{t_5:50}E$. From special case 2, the temporal cycles starting from edge $A\xrightarrow{t_1:100}C$ gets highest priority while the temporal cycles starting from edge $A\xrightarrow{t_6:50}C$ gets the least priority. Next at node C, there are three outgoing edges but considering case 3 only the edge  $C\xrightarrow{t_4:50}D$ gets priority. Similarly, now at node D, edge from node D to node A gets priority. The reason for not considering the other edge $D\xrightarrow{t_5:50}E$ in the temporal cyclic path starting from A is due to our assumption that no other node except starting node is allowed to repeat more than once in a cycle. Using special case 1, edges from node C to node D and from node D to node A get merged. So the first three temporal cycles results in only one temporal cycle (cf. Table~\ref{tab:cyclesList}). After applying the different cases, the fourth cycle becomes invalid because we do not allow the repetition of edges in any temporal cycles.

\begin{algorithm}
\caption{Temporal Cycle Detection Method}\label{alg:allCycles}
\begin{algorithmic}[1]
\State \textbf{Input}
\State Blockchain's Transaction Graph ($G\gets (V,E)$)
\State visVertex$\gets \Phi$ \{Two states: visited and unvisited\}
\State visEdge $\gets \Phi$ \{Three states: finally visited, partially visited and unvisited\}
\State ~
\State orderedEdges $\gets \Phi$
\For {account $\in$ startingAccounts}
    \For {edge $\in G[account]$}
        \State orderedEdges.append($<edge>$)
    \EndFor
\EndFor
\State orderedEdges.sort() \#based on $<blocknumber,valueTransferred,gasPaid>$
\For {edge $\in$ orderedEdges}
    \State $previousPath\gets \{\}$
    \State AllCycles(previousPath.append($<edge>$))

\EndFor
\State ~
\Procedure{AllCycles}{previousPath=$S\xrightarrow{t_0:x_0}v_0\xrightarrow{t_1:x_1}v_1\xrightarrow{t_2:x_2}v_2\cdots\xrightarrow{t_k:x_k}v_k$}
\State $currAcc\gets v_k$
\State $currBlockNo\gets t_k$
\State $flag\gets 0$

\If {$S ==$ currAcc} 
    \State \textbf{OUTPUT} previousPath.append$<(currAcc,S,t)>$
    \State Mark all the edges involved in this cycle as finally visited
    \State \textbf{return} True
\EndIf

\State $outGoingEdges\gets \{(currAcc,v,t)\in E$ and $currBlockNo <t \}$
\State outValue$\gets 0$

\For {temp $\in$ outGoingEdges } 
    \State flag1$\gets $0
    \State tmin $\gets$ temp.blockNumber
    \If{tmin>currBlockNo and \textbf{visEdge[temp]} and \textbf{visVertex[temp.to()]} are unvisited}
        \If{$outValue<inValue$}
            \State visEdge[temp] $\gets$ Mark partially visited
            \State visVertex[currAcc] $\gets$ Mark Visited
            \State flag1$\gets$ AllCycles(previousPath.$<(currAcc,temp,t_{min})>$)
            \If{flag1==True}
                \State flag $\gets$True
                \State outValue+=temp.value
            \EndIf
        \EndIf
    \EndIf
\EndFor
\State visVertex[currAcc]$\gets$ Mark unvisited
\State \textbf{return} $flag==True$
\EndProcedure
\end{algorithmic}
\end{algorithm}

\begin{figure}
    \centering
    \subfloat[Example 1]{
        \includegraphics[width=0.4\textwidth]{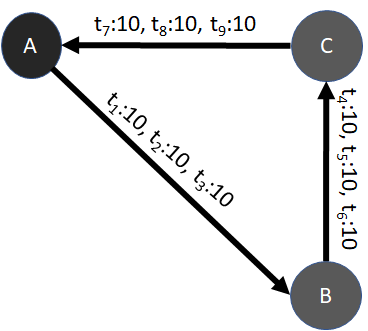}
        \label{fig:exapmle1}
    }
    \subfloat[Example 2]{
        \includegraphics[width=0.4\textwidth]{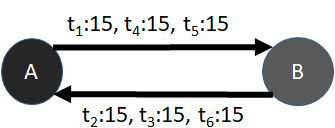}
        \label{fig:example2}
    }
     \caption{Examples to show that how modifications can be applied for reducing the number of temporal cycles}
    \label{fig:usefullnessOfAssumptions}
\end{figure}

As another example, in Figure~\ref{fig:exapmle1}, without using our modifications, there are, in total, 27 cycles, but after applying the first modification, the total number of temporal cycles reduce to only 1. Total money transferred along the temporal cycle after applying the modifications is equal to the total money (30 cryptocurrencies) transferred before applying our modifications. Thus, using our modifications, there is no loss in the money. Similarly, in Figure~\ref{fig:example2}, node A and node B transact with each other. Here, the total number of temporal cycles is 9 (5 starting from node A and 4 starting from node B). After applying the second and the third modification, only 3 temporal cycles remain. 

\subsection{Our Algorithm for finding Temporal Cycles}

A depth-first search (DFS) algorithm is a well-known algorithm used to detect cycles in a static graph. For each connected component, DFS produces a DFS tree. A cycle is present in the graph if there is a back edge (an edge that joins a node to itself or one of its ancestors in the DFS tree). We use a modified DFS algorithm based on the cases described above to find temporal cycles in the graph (cf. Algorithm~\ref{alg:allCycles}). Following are the inputs to our algorithm:

\begin{itemize}

\item [(a)] The transactions graph $G(V, E)$ represents graph generated using the transactions between nodes over time. Here, $V$ is the set of user accounts in the blockchain. $E$ is the set of directed transactions between 2 accounts in $V$. Note that these transactions are temporal. We use the block number to represent the time at which the transaction occur. An edge (or a transaction) in $E$, besides timestamp, also contains information related to the transaction such as value (money) transferred, gas paid for the transaction, and whether it is an internal transaction or not. 

\item [(b)] Consider $startingAccounts$ to be a set of all malicious accounts. From $S\in startingAccounts$ our temporal cycles starts because we aim to analyze different tagged malicious activities.

\item [(c)] We use a depth-first search (DFS) based recursive approach starting from node $S\in startingAccounts$ to find cycles. A path to a node $v_k$ from $S$ is called the $previousPath$. The $previousPath$ could be understood as a fragment of a potential cycle. Note that $v_k$ represents current account from which next transactions is to be explored for next edge in the temporal path. An example of $previousPath$ is $S\xrightarrow{t_0:x_0}v_0\xrightarrow{t_1:x_1}v_1\xrightarrow{t_2:x_2}v_2\cdots\xrightarrow{t_k:x_k}v_k$.


\item [(e)] There are three types of edges, (\textit{i}) visited edges (edges that belong to a cycle), (\textit{ii}) contenders (edges that are in a temporal path which may evolve to a cycle), and (\textit{iii}) not visited edges (edges that are not yet considered). $visEdge$ mark the set of edges that already belong to all the temporal cycles that were previously explored. Edges that are involved in the $previousPath$ are the contenders. $visEdge$ is used to know if an edge has already occurred in any of the detected temporal cycles. 

\item [(f)] Similar to $visEdge$, $visVertex$ is used to know whether a vertex is already visited or not for a particular cycle. Note that a vertex can be in multiple cycles. Thus, $visVertex$ is local to each cycle.

\item [(e)] We sort all the transactions from all $S\in startingAccounts$ and store them in the $orderedEdges$. First sorting is based on the timestamps (at which the transactions occur), the lower timestamp of a transaction higher priority it gets. If timestamps are same for two transactions, then we sort based on higher cryptocurrency transferred. If the cryptocurrency transferred is also same for the two transactions, then we prioritize the transaction based on other attributes in $x$ such as gas price (transaction that pays more gas).
\end{itemize} 

Our recursive approach returns all the temporal cycles. We initialize the $previousPath$ with the edge from which temporal cycles start (an edge from $orderedEdges$) and node $S\in startingAccount$. During the recursive exploration of the temporal graph in the depth-first search manner, starting from the last visited account in the previous path, there is a temporal cycle if the path ends at $S$. 
At the current node, we explore all outgoing edges in sorted order (cf. Algorithm~\ref{alg:allCycles} line 25), which respect temporal aspects of the temporal path defined in $previousPath$. At each step, if the sum of transacted money (value transferred in a transaction) in the outgoing edges involved in the temporal cycle is greater than incoming money at that node from the incoming edge (cf. Algorithm~\ref{alg:allCycles} line 29), then further exploration of the outgoing edges at that node is stopped and the node is marked visited (cf. Algorithm~\ref{alg:allCycles} lines 31-32). Not allowing further exploration is based on special case 3 because once we trail the complete money for an incoming transaction at any node, we stop to trail that money further in the graph. When a cycle is detected (cf. Algorithm~\ref{alg:allCycles} lines 20-22), all the edges involved in the temporal cycle are marked as visited (one edge is allowed to appear in only one temporal cycle). A visited edge during exploration of a particular temporal cycle is not revisited. The procedure \textbf{AllCycles} returns true if there is a temporal cycle, else it returns false. Using the procedure and steps defined above, we list all the valid temporal cycles based on our modifications. 

Our modifications does not change the functional approach of the DFS algorithm. As shown in~\cite{2scent,simpleCycles}, the DFS-based approach is valid for finding temporal cycles of n-hop length where $n\in \mathbf{N}$. The time complexity of our algorithm is dependent on the number of temporal cycles in the temporal graph. The time complexity of the DFS approach is $\mathcal{O}(|V|+|E|)$. Thus, the time complexity of our approach is $\mathcal{O}(C*(|V|+|E|))$, where $C$ is the number of valid temporal cycles present in the transaction graph. 

\section{Evaluation}\label{sec:eval}

This section first presents the details of the data we collect and the pre-processing we perform. We then provide a detailed analysis of our results based on our methodology.

\subsection{DataSet}\label{sec:dataset}

There are two most prominent cryptocurrency-based permissionless blockchains: Bitcoin~\cite{bitcoin} and Ethereum~\cite{ethereum}. We choose Ethereum blockchain transaction data because it is more rich and diverse. Etherscan~\cite{etherscan} makes available the information about accounts and transactions categorised into different illicit activities. 

Ethereum has two types of accounts Externally Owned Accounts (EOA)s and Smart Contracts (SC)s. SCs get executed on Ethereum Virtual Machine (EVM). The transactions between two SCs (also called internal transactions) are not stored on the ledger, but they can be inferred using EVM. Transactions between two EOAs and EOAs and SCs are stored on the ledger of the Ethereum blockchain and are called external transactions. These transactions are publicly available using Etherscan APIs~\cite{etherscan}. We, thus, use the Etherscan APIs to download the transactions. Both the types of transactions have a different internal JSON structure. For our analysis, we consider all the internal and external transactions of the chosen accounts from the genesis block (block number 0) to block number 10747845 (generated on date: 28/08/2020).

\begin{figure}
\includegraphics[width=0.8\textwidth]{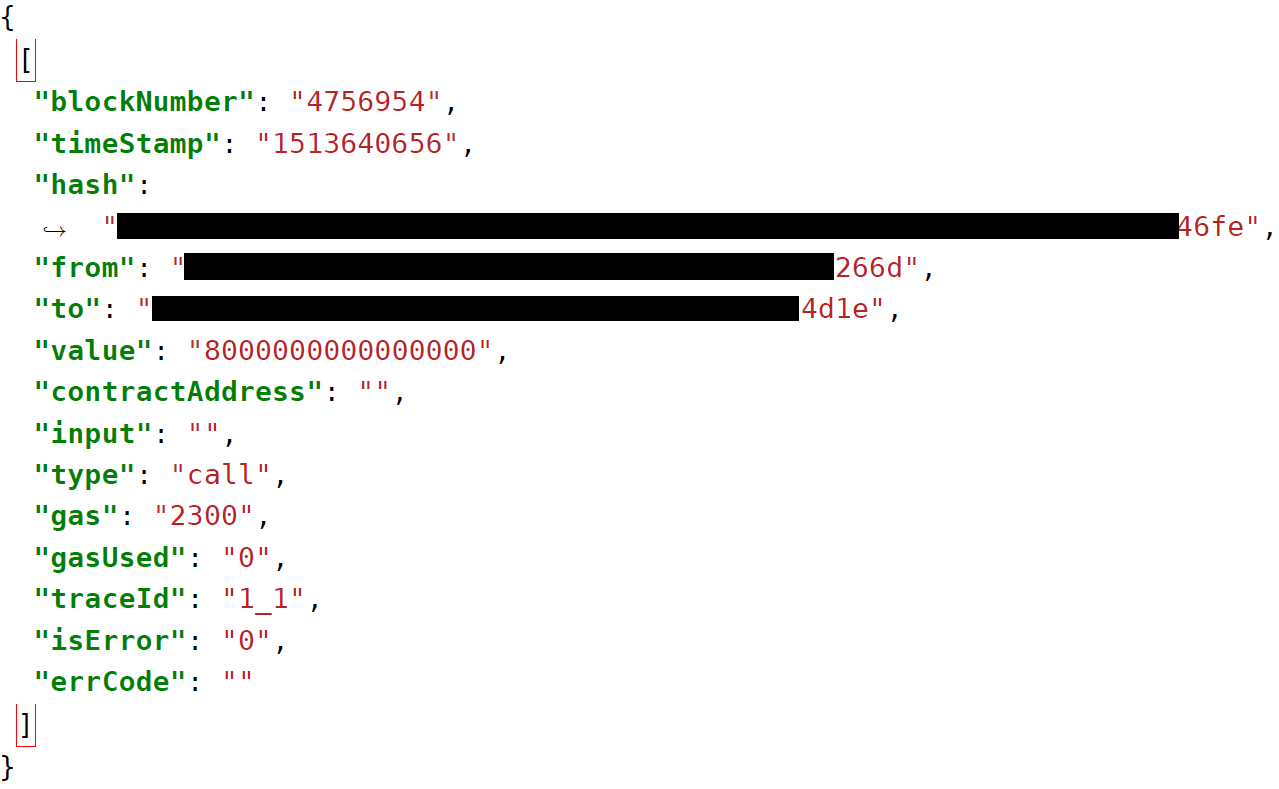}
\caption{Sample Internal Transaction in Ethereum}
\label{fig:internalAccount}
\end{figure}

\begin{figure}
\includegraphics[width=0.8\textwidth]{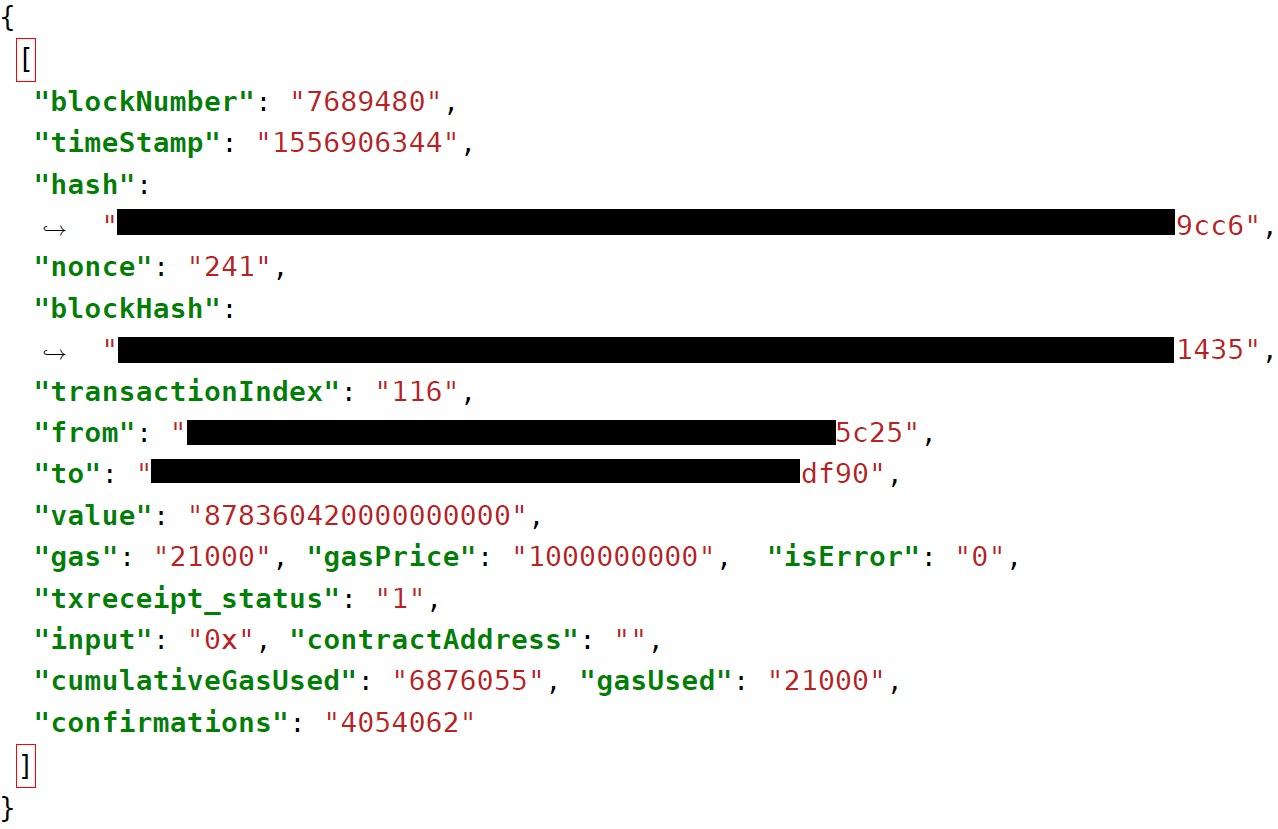}
\caption{Sample External Transaction in Ethereum}
\label{fig:externalAccount}
\end{figure}

\subsubsection{Transaction Structure}

As described before, the two types of transactions have different structures.
A sample internal transaction shown in Figure~\ref{fig:internalAccount} while a sample external transaction is shown in Figure~\ref{fig:externalAccount}. We now describe the terminology used in these types of transactions and only those that help us understand the money trails.

\begin{itemize}
    \item \textbf{from:} hash of the sender address of the transaction.
    \item \textbf{to:} hash of the receiver address of the transaction.
    \item \textbf{hash:} transaction hash.
    \item \textbf{blockHash:} hash of the block in which transaction appears.
    \item \textbf{blockNumber:} it is the block to which transactions appears.
    \item \textbf{timeStamp:} milliseconds after epoch when the block was generated.
    \item \textbf{value:} It is the amount of cryptocurrency transferred in Wie. For Ethereum, $10^{18}$ Wei equals 1 Ether.
    \item \textbf{contractAddress:} hash of a contract address. It is not null only if the transaction is a contract create transaction.
    \item \textbf{gas:} The maximum amount of gas units that the transaction can consume. Units of gas represent computational steps.
    \item \textbf{gasUsed:} The amount of gas used for a transaction.
    \item \textbf{gasPrice:} the price of Gas in Gwei. For Ethereum, $10^{9}$ Gwei equals 1 Ether.
    \item \textbf{isError/txreceipt\_status:} True if transaction is unsuccessful else false.
\end{itemize}

\subsubsection{Dataset Statistics}

Some of the most common malicious activities involved in the blockchain are Gambling, Phishing, and Money Laundering. These activities have resulted in a sizable losses of cryptocurrency over time in the blockchain. With the increasing popularity of blockchain, adding new dimensions to the analysis of these activities in the blockchain is necessary. Also, in Ethereum, these activities have enough tagged accounts and transactions to understand their behavior in terms of money trails. 

Table~\ref{table:stats} details the statistics about the transaction dataset we extract from Ethereum using Etherscan APIs~\cite{etherscan} for our methodology validation. In the Table~\ref{table:stats} we present two types of statistics. One, representing that which shows the statistics  where we present the EOAs, SCs, and transactions between accounts of specified malicious activity. In the second, we present statistics where we present the EOAs, SCs, and transactions between the accounts of specified malicious activities and their one-hop neighbors. Note that here the one-hop neighbors include benign accounts as well. Our results presented next consider these two types. For simplicity, from here, we refer to these two types as \textit{type-A} and \textit{type-B} and demonstrate the results obtained using transactions involved in these two types.

In~\cite{agarwal2021detecting}, the authors used more (such as Heist, Scamming) malicious activities other than those we consider. Other than Scamming malicious activities, all other malicious activities have fewer tagged accounts and transactions. We categorize these activities as one and under the class ``other'' malicious activities. Most of the scamming-related accounts are also tagged as Phishing in Ethereum. Thus, we consider these accounts only once under Phishing. For ``other'' malicious activities (total 315 accounts), we have used data only to analyze whether different types of malicious activities transfer money in the temporal cyclic path with each other or not. Thus, we have not included the information for other malicious activities in Table~\ref{table:stats}.

\begin{table}
\caption{Statistics}
\label{table:stats}
\small
\begin{tabular}{|c|c|ccc|}
\hline 
     & \textbf{including one-hop} & &  & \textbf{Total}\\
    \textbf{Type of Activity} & \textbf{benign neighbors} & \textbf{EOAs} & \textbf{SCs} & \textbf{Transactions}\\
    \hline
    \multirow{2}{*}{Gambling} & \xmark &  4 & 38 & 930 \\
    \cline{2-5}
    & \cmark & 33158 & 67289 & 5942672 \\
    \hline
    \multirow{2}{*}{Phishing} &\xmark & 4076 & 693 & 1422 \\
    \cline{2-5}
    & \cmark & 24527 &5893 & 1108836 \\
    \hline
    \multirow{2}{*}{Money Laundering} & \xmark & 813 & 2 & 2157 \\
    \cline{2-5}
    & \cmark & 11728 & 3021 & 256113\\
    \hline
   \end{tabular}
\end{table}

\subsection{Data Pre-processing}

\begin{figure}
    \centering
    \subfloat[More than one transaction between two accounts in the same block]{
        \includegraphics[width=0.4\textwidth]{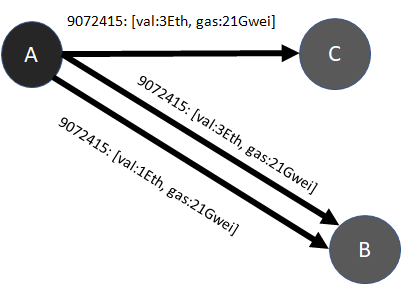}
        \label{fig:sameBlockTransaction}
    }
    \subfloat[Merged Transactions of Same Block ]{
        \includegraphics[width=0.4\textwidth]{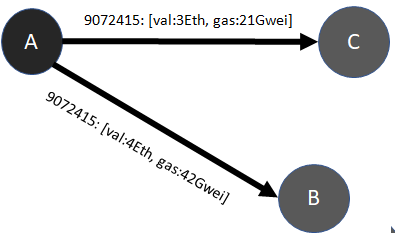}
        \label{fig:merged}
    }
     \caption{Processing of transactions happening between two accounts in the same block}
    \label{fig:transactionPreProcessing}
\end{figure}

In our experiments, we use only successful internal and external transactions. From the internal and external transaction data, we construct a graph that contains the details such as (\textit{a}) from, (\textit{b}) to, (\textit{c}) block number, (\textit{d}) isInternalTransaction: true if the transaction is internal else false, (\textit{e}) value, (\textit{f}) gas, and (\textit{g}) gas used. We use block numbers to represent timestamp because we do not have the exact time when the transaction happened. We remove the transactions which transfer 0 Eth (Cryptocurrency used by Ethereum). Transactions that have 0 Eth transferred can give some important insights. But in our work, we have not used such transactions because these transactions result in 100$\%$ money loss, but we aim to find the suspicious cycles with very little loss along the path. In the future, these transactions also can be included. Each transaction represents a directed temporal edge in the graph, except if account A sends money to an account B multiple times in the same block, then we merge these transactions into a single transaction. As shown in Figure~\ref{fig:sameBlockTransaction}, account A transacts with account B two times in the same block represented by block number 9272415. Thus, we merge both of these transactions into one transaction (cf. Figure~\ref{fig:merged}). With the computational resources currently available to us, it is impossible to perform experiments on the whole Ethereum blockchain for path-based analysis of the graph due to the large size of the Ethereum blockchain. We perform our experiments on the tagged malicious accounts (particularly Phishing, Gambling and Money laundering). One type of Money laundering activity in the Ethereum blockchain is the activity (commonly called as UpBit hack) that involved the UpBit exchange. There are speculations that some insider performed malicious activity during the movement of money from a hot to a cold wallet~\cite{upBit}). In this work, we use accounts tagged as ``Upbit Hack'' in the Ethereum blockchain as Money laundering accounts. 

\subsection{Results} \label{sec:results}
This section presents results obtained. The presented results pertain to when  accounts involved in different types of malicious activities: Gambling, Phishing, Money Laundering, other malicious activities, as well as when  considering all malicious activities under one class.

\subsubsection{Gambling}

We first apply our methodology to accounts labeled under the Gambling category by Etherscan~\cite{etherscan}. As a first step, we generate the graph from these accounts and their transactions. Here, we consider only those transactions which happen between only Gambling accounts. As it is usually done, for cryptocurrency also, we assume that accounts associated with Gambling activity put their money in a depository (or an exchange in terms of blockchain) for betting/lottery. Upon win, the account gets back the funds assured (for example, people Gambling in casinos~\cite{casino}). 
Using our approach, we find that out of the 42 marked Gambling accounts (4 EOAs and 38 SCs), only 2 SCs have temporal cycles, that too cycles involving only themselves. We do not disclose the addresses of these accounts  due to privacy reasons. However, one of these  accounts is an exchange. 
We find that the total number of temporal cycles is 136958, if we do not apply our modifications defined in Section~\ref{sec:method}. After the application of our modifications, the total number of temporal cycles reduces to 346. All of these temporal cycles are of two-hop length. Note that two-hop length does not mean that the transactions happen in two consecutive time instances. The inter-event time (time between the two transactions) can be $>1$. The two accounts involved in the temporal cycles transact with each other regularly. All the transactions involved in the temporal cycles are internal transactions as the transactions happen between two SCs. From these results, it is clear that most Gambling accounts are not transacting with each other for Gambling activities. However, the accounts involved in the temporal cycle follow a traditional gambling approach.

Next, we add more accounts to extract more insights and know whether our assumptions hold when accounts not marked as Gambling are added. Here, we add those accounts to the graph that are not marked as Gambling but have received funds from the marked Gambling accounts, i.e., Gambling accounts have an out edge to these accounts. The added accounts have high cosine similarity (calculated using features defined in~\cite{agarwal2021detecting}), more than 0.99, with the tagged Gambling accounts. The number of temporal cycles does not change after adding these accounts and their transactions (limited to considered accounts). Next, we add the neighbors (neighbors of neighbors of Gambling accounts that we added in the last step). We repeat adding neighbor accounts to the graph four times. But, in all the steps, the number of temporal cycles does not change. It means that most Gambling accounts do not show temporal cyclic behavior with other Gambling accounts or accounts transacting directly to marked Gambling accounts.

\begin{figure}
    \centering
    \subfloat[Example 1 ]{
        \includegraphics[width=0.4\textwidth]{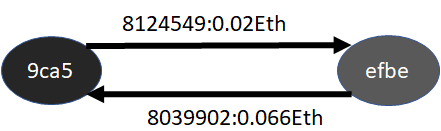}
        \label{fig:gambleExample1}
    }
    \subfloat[Example 2 ]{
        \includegraphics[width=0.4\textwidth]{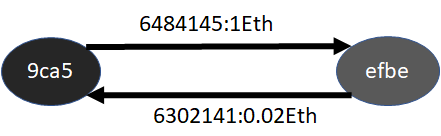}
        \label{fig:gambleExample2}
    }
     \caption{Suspicious Temporal Cycles in the Gambling}
    \label{fig:gambling}
\end{figure}

To identify the largest cycle, we add all the accounts (and their transactions) that have a path from the Gambling accounts into our graph. We find that the number of temporal cycles increases many-fold ($>10^{14}$). We find that 6 accounts contribute the most number of temporal cycles in Gambling. With the limited computing  resources we have had at our disposal, we were not able to extract the exact number of cycles from these 6 Gambling accounts. For the remaining 36 Gambling accounts, we find that the total number of cycles is 1313104, and the maximum hop length of the temporal cyclic path is 23. We find that these 36 accounts are involved in temporal cycles that do not include exchange accounts. These results show that the Gambling accounts do not transfer money in a temporal cyclic path to other Gambling accounts while the Gambling accounts have money trails with other non-gambling accounts. We find few suspicious temporal cyclic path-based money transfers (cf. Figure~\ref{fig:gambling}). In Figure~\ref{fig:gambling}, we represent a node by the last four characters of the account address hash and an edge by the key-value pair where the key is the block number in which the transaction appears. The value represents the amount of Ether transferred via the transaction. Money loss along the temporal cycles present in Figure~\ref{fig:gambling} is negative. This type of behavior indicates that there is a possibility that the account won the bet/lottery. Our approach gives all of the temporal cycles for a given threshold of maximum money loss allowed along the cyclic path. For money loss $>$10\%, we find that the number of temporal cycles does not increase significantly, while the number of cycles with money loss  $<$10\% is significantly high in the Gambling.  

\subsubsection{Phishing}

\begin{figure}
    \centering
    \subfloat[Example 1 ]{
        \includegraphics[width=0.4\textwidth]{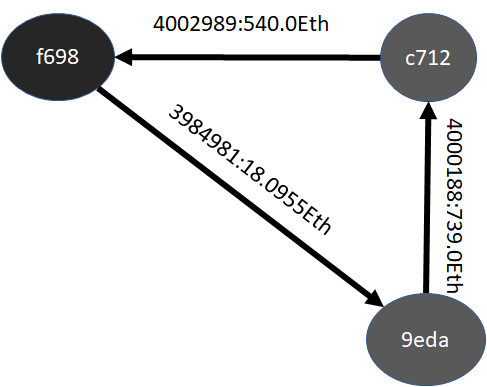}
        \label{fig:phishExample1}
    }
    \subfloat[Example 2 ]{
        \includegraphics[width=0.4\textwidth]{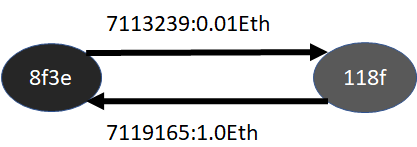}
        \label{fig:phishExample2}
    }\\
     \caption{Some of the Suspicious Temporal Cycles in the Phishing }
    \label{fig:phishing}
\end{figure}

We now apply our methodology to accounts labeled under Phishing. There are a total of 4769 Phishing tagged accounts (4076 EOAs and 693 SCs). We find that a total of 103 accounts are involved in the time respecting cyclic transfers. There are 55 accounts from which a minimum of one cycle starts. Most Phishing accounts have very few transactions. This limits the chances of them being present in a temporal cycle. There are a total of 1682 cycles in the graph before applying our modifications. After the application of our modifications, the total number of temporal cycles reduce to 164. Here, most of the cycles are of two-hop. The maximum hop-length of a cycle obtained is 3. We find that some of these transfers show suspicious behavior. As shown in Figure~\ref{fig:phishing}, money loss along the cyclic path is less than 10\%.

Then, similar to the analysis of Gambling accounts, we add out-neighbors (accounts to which Phishing accounts send funds) of Phishing accounts and have high similarities ($>0.99$) with them. We observe that the number of cycles does not change. After that, we add more accounts to get more insights and know whether the results are consistent when the graph size increases. Irrespective of the similarity, we add all the out-neighbors of the tagged Phishing accounts. We find $>10^8$ temporal cycles, a significantly high number when our modifications are not applied. However, only 4315 cycles remain when we apply our modifications. In these 4315 cycles, the maximum hop-length is 20. Figure~\ref{fig:phishExample1} shows one suspicious cycle detected in the analysis of Phishing accounts. Here, the money loss along the cyclic path is negative, meaning Phishing accounts have received funds. Also, 1945 temporal cycles have less than 10\% money loss. 

\subsubsection{Money Laundering}

Next, we study money laundering-based accounts. As it traditionally happens, for cryptocurrency-based blockchains, we assume that accounts involved in money laundering will produce more cycles than those involved in Gambling or Phishing. For a special case of Money laundering, depicted by accounts related to Upbit exchange, there are 815 tagged accounts. Here, out of these 815 accounts, only two accounts are SCs. We extract both internal and external transactions for these accounts. However, all the internal transactions have an error flag `=True' (meaning the transaction is unsuccessful). Thus, we do not consider these transactions. Most of these accounts have very few transactions (3 to 5 transactions only). Thus, the chances of these accounts appearing in a temporal path-based cyclic path is less. 

\begin{figure}
    \centering
    \subfloat[Example 1 ]{
        \includegraphics[width=0.4\textwidth]{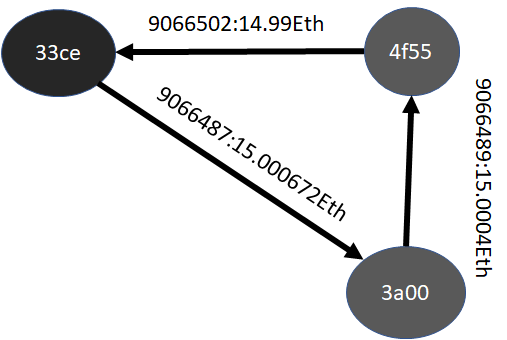}
        \label{fig:moneyExample1}
    }
    \subfloat[Example 2 ]{
        \includegraphics[width=0.4\textwidth]{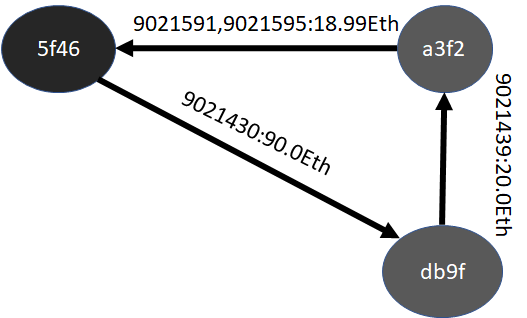}
        \label{fig:moneyExample2}
    }\\
    \subfloat[Example 3 ]{
        \includegraphics[width=0.4\textwidth]{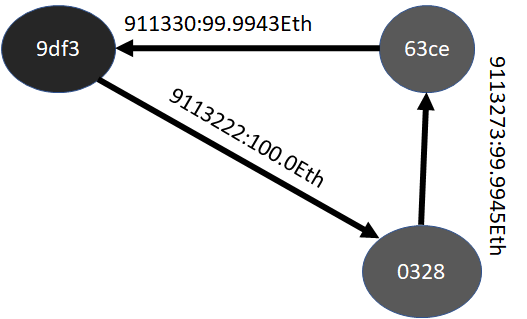}
        \label{fig:moneyExample3}
    }\\
     \caption{Some of the Suspicious Temporal Cycles in the Money Laundering }
    \label{fig:moneyLaunderingWithoutFirstHop}
\end{figure}

\begin{table}
\caption{Results}
\label{table:results1}
\begin{tabular}{|c|c|c|c|c|c|c|}
\hline
    \textbf{Activity} & \multicolumn{2}{|c|}{Gambling} & \multicolumn{2}{|c|}{Phishing} & \multicolumn{2}{|c|}{Money Laundering} \\
    \hline
    \textbf{Considering} & type-A & type-B & type-A & type-B & type-A & type-B \\
    \hline
    \textbf{No. of cycles} & 346 &131304 &164 & 4315 &40 &90  \\
    \hline
    \textbf{Max. Hop Length} &  2 &23 &3 & 20 &6 &6  \\
    \hline
    \textbf{No of cycles with}& \multirow{2}{*}{337} &\multirow{2}{*}{50108} &\multirow{2}{*}{128} & \multirow{2}{*}{1945} &\multirow{2}{*}{8} &\multirow{2}{*}{42}  \\
    \textbf{($\alpha\leq10\%$) loss } &&&&&& \\
    \hline
    \textbf{No. of unique accounts} & \multirow{2}{*}{2} & \multirow{2}{*}{4305} & \multirow{2}{*}{103} & \multirow{2}{*}{3196} & \multirow{2}{*}{69} & \multirow{2}{*}{150} \\
    \textbf{involved in all cycles} &&&&&&  \\
    \hline
\end{tabular}
\vspace{-0.4cm}
\end{table}

As before, we first analyze the transactions between only the tagged accounts. We find that there are a total of 83 temporal cycles when we do not apply our modifications. After the application of our modifications, only 40 temporal cycles remain. Here, 69 unique tagged accounts are involved in these temporal cycles. The maximum hop length in these temporal cyclic paths is 6. It means money laundering-based accounts are more actively involved in the cyclic temporal path-based money transfers. Thereby validating that money laundering based accounts transact in cycles in cryptocurrency-based blockchain as well. Figure~\ref{fig:moneyLaunderingWithoutFirstHop} shows a suspicious temporal cyclic path-based transfer. As before, the node name represents the last four characters of the tagged malicious accounts in the Ethereum blockchain. Loss of the money along the temporal cyclic paths in Figure~\ref{fig:moneyExample1} is significantly less. In Figure~\ref{fig:moneyExample1}, the completion time for cyclic transfer is 15 (starting block number is 9066502 while the ending block number is 9066487) blocks. This is equivalent to $\approx$3min in the Ethereum blockchain. Money transfer along the temporal cyclic path in a short span indicates illicit activity. But the money lost in Figure\ref{fig:moneyExample2} is high. Note that our graph is only limited to Money laundering accounts. It is quite possible that adding all of the neighbors and their neighbors continuously until no new neighbor is found will  result in the recovery of the whole amount in the cyclic path-based transfer. In Figure~\ref{fig:moneyExample2}, despite high money loss except for the first transaction, all other transactions have very little loss of money along the path. One of the reasons for these types of cycles is the limited data set. We add all the out-neighbors of the money laundering accounts into our graph and their corresponding transactions to get more consistent results. If we consider our modifications, the number of cycles we get is only 90. The maximum number of hops present in these cycles is 6. The number of cycles has not increased significantly, but the cycles with very less loss (less than 10\% money loss) during cyclic temporal path-based transfer increased.  The temporal cycles in which high loss occurs indicate suspicious activity. Figure~\ref{fig:moneyExample3} shows one of the detected suspicious temporal cyclic path-based transfer in money laundering after adding one-hop neighbors. 

In short, our results are summarized in Tables~\ref{table:results1} and~\ref{table:results2}. Table~\ref{table:results1} provides results for the number of valid temporal cycles, maximum hop-lengths in the cycles found using our approach for a particular malicious activity, and the number of temporal cycles in which money loss along the temporal cyclic path is less than 10\%. While Table~\ref{table:results1} lists the number of accounts that are involved in any valid temporal cycles. Table~\ref{table:results2} presents an exact number of cycles for  valid temporal cycles of different hop-lengths  as found using our approach for the  different malicious activities.

\begin{table}
\caption{Distribution of cycles concerning Hop-length obtained using transactions involved in the two types defined previously.}
\label{table:results2}
\centering
\begin{tabular}{|c|c|c|c|c|c|c|}
\hline
     & \multicolumn{2}{|c|}{Gambling} & \multicolumn{2}{|c|}{Phishing} & \multicolumn{2}{|c|}{Money Laundering}   \\
     \cline{2-7}
     Hop-Length &type-A & type-B & type-A & type-B & type-A & type-B \\
     \hline
     2 & 346 & 113136 & 156 & 3812 & 29&54  \\
     \hline
     3 & 0 & 4718 & 8 & 246 & 6 & 14  \\
     \hline
     4 & 0 & 12087 & 0 & 135 & 2 & 5 \\
     \hline
     5 & 0 & 826 & 0 & 40 & 2 & 16 \\
     \hline
     6 & 0 & 334 & 0 & 23 & 1 & 1 \\
     \hline
     7 & 0 & 81 & 0 & 27 & 0 & 0 \\
     \hline
     8 & 0 & 63 & 0 & 11 & 0 & 0 \\
     \hline
     9 & 0 & 8 & 0 & 4 & 0 & 0 \\
     \hline
     10 & 0 & 17 & 0 & 2 & 0 & 0 \\
     \hline
     11 & 0 & 8 & 0 & 5 & 0 & 0 \\
     \hline
     12 & 0 & 6 & 0 & 5 & 0 & 0 \\
     \hline
     13 & 0 & 9 & 0 & 1 & 0 & 0 \\
     \hline
     14 & 0 & 0 & 0 & 0 & 0 & 0 \\
     \hline
     15 & 0 & 0 & 0 & 2 & 0 & 0 \\
     \hline
     16 & 0 & 0 & 0 & 1 & 0 & 0 \\
     \hline
     17 & 0 & 8 & 0 & 0 & 0 & 0 \\
     \hline
     18 & 0 & 1 & 0 & 0 & 0 & 0 \\
     \hline
     19 & 0 & 0 & 0 & 0 & 0 & 0 \\
     \hline
     20 & 0 & 0 & 0 & 1 & 0 & 0 \\
     \hline
     23 & 0 & 2 & 0 & 0 & 0 & 0 \\
     \hline
\end{tabular}
\vspace{-0.4cm}
\end{table}

\subsubsection{Other Activities}

We also apply our approach to other malicious activities. We find that 113 out of 214 scamming accounts are tagged as Phishing accounts. We find that these scamming accounts are not involved in any cyclic path-based money transfer. Even after adding one-hop neighbor accounts, which are tagged both as scamming and Phishing, we do not find any temporal cycles originating except from two EOAs. Thus, it is clear that the accounts that are tagged both as Phishing and scamming behave differently from the other Phishing accounts and that the Phishing accounts do cluster themselves into more than one cluster. 

\subsubsection{Combined Graph}

Next, we combine accounts of all the malicious activities (including ``other activities'') to know whether malicious activities in the blockchain do cyclic path-based money transfers with other types of malicious activities. We find that only one Gambling account is involved in a cyclic transfer with two Phishing accounts. But the Gambling account involved in the cyclic transfer is an exchange account. Accounts involved in malicious activities in the Ethereum blockchain do not transfer cryptocurrency in a cyclic path to accounts involved in other malicious activities. Also, we found that ``other'' malicious activities have no cyclic path-based money transfer with themselves except for 3 accounts. Thus these ''other'' activities also cluster themselves into a different cluster.

From our results, it is clear that money laundering-based malicious activities are different from other malicious activities. Phishing-based accounts do not cluster into one cluster. Phishing accounts cluster themselves into a minimum of two different clusters (accounts also involved in scamming do cluster into different clusters than the accounts that are only involved in Phishing activity). Also, most of the Gambling accounts do not transfer money in a cyclic path with themselves. After adding more transactions into validation data, we find that Gambling accounts are transacting with neighbors regularly. Also, Gambling and Phishing-based accounts that are not involved in scamming show a similar behavior to Phishing accounts after adding neighbor accounts. Both Gambling and Phishing activities transfer money using temporal cycles having larger hop counts, and the number of temporal cycles also increases. With the availability of limited computing  resources, we could not validate our methodology on a larger dataset. Nonetheless, our approach can be extended to a larger dataset where more computation power is available. 

From our experiments, it is clear that most malicious activities can be clustered into four clusters. One cluster has money laundering-based activity, another cluster has Phishing and Gambling-based activities, the third cluster contains scamming and Phishing-based activity, and the fourth cluster contains ''other'' malicious activities. Using Neural Networks, in~\cite{agarwal2021detecting}, the authors showed that most of the marked malicious activities in the Ethereum blockchain clusters into four different clusters, but they have not given in-depth insights on why such type of similarities and differences occur. This work also obtains similar clustering indications by exploring a hidden aspects in the blockchains transaction network, i.e., money-trails. 

\section{Conclusion}\label{sec:conclusion}

With the evolution of blockchain 2.0 and beyond, interests  in blockchain technology is on a rapid rise. New use cases (such as healthcare for making a secure system for health records) of blockchain have created many applications worldwide that neither support any cryptocurrency nor crypto-tokens. As more and more solutions based on  blockchain technology are introduced, we find more cases of exploitation of the technology and  its  users. Most of these  exploit previously unknown vulnerabilities in the technology stack or gullibility of the users to fall for social engineering attacks. Such attacks result in substantial disruptions and losses of digital  assets. In permission-less public blockchains such as cryptocurrency blockchains, the losses amount to monetary loss. Over the last few years,  various state-of-the-art approaches have been proposed  to counter attacks and detect suspicious behavior in the blockchain. Most of these approaches are  ML-based and do not distinguish between multiple classes of malicious activities and club them together under one class. While others focus only on specific kind of malicious activity. 
Further, these techniques do not track the cryptocurrency flow. 

In this work,  we use a 'track the money trail' approach for collecting additional behavioral information regarding  different malicious activities. We use temporal transactions network for finding temporal cycles to track the money flow  along these temporal cyclic paths. Based on the temporal cyclic path and money loss results along the path, we found that the considered malicious activities are clustered into four clusters. Also, we found that accounts involved in the Phishing activities cluster themselves into more than one cluster. We found that accounts involved in both Phishing and scamming cluster into different clusters than other accounts involved in  Phishing only. Thus, it bears out that  suspicious cyclic path-based money transfers in cryptocurrency blockchains  can distinguish suspicious accounts of various types of suspicious activities. 

Illicit activity detection in a blockchain is  challenging yet much  necessary. There is a demand for forensics that involves finding suspicious activities and attribution from law enforcement agencies, regulators, and regular participants in cryptocurrency blockchains. In future, we aim to integrate our approach with the machine learning approaches and use the properties of these money-trail  cycles as features. This may help to detect suspicious activities with high accuracy. Also, parallel computing can be used to reduce computation time for the cycle identification algorithms.  We have used only EOAs and SCs based transactions; in the future, the token transfers paths and cycles may also be used to get more insights through  our approach.

\section*{List of Abbreviations}
\begin{table}[ht]
\centering
\begin{tabular}{|c|l|}
\hline
\textbf{Acronym} & \textbf{Meaning} \\
\hline
EOA & Externally Owned Accounts\\
SC & Smart Contracts \\
Blockchain & Permission-less Blockchain \\
ICO & Initial Coin Offerings \\
DFS & Depth First Search \\
FH & First Hop Outgoing Neighbours of an Account in the Transaction Graph \\
ML & Machine Learning \\
NN & Neural Network \\
SVM & Support Vector Machine\\
CCG & Contract Creation Graph \\
CIG & Contract Invocation Graph \\
MFG & Money Flow Graph \\
ATH & Attributed temporal Heterogeneous Motif \\
AAIN & Homogeneous Address-Address Interaction Network \\
TAIN & Heterogeneous Transaction-Address Interaction Network\\
\hline
\end{tabular}
\caption{List of Acronyms.}
\label{tab:acronym}
\end{table}





  
\section*{Acknowledgements}
This work is partially funded by the National Blockchain Project at IIT Kanpur sponsored by the National Cyber Security Coordinator's office of the Government of India and partially by the C3i Center funding from the Science and Engineering Research Board of the Government of India.

\bibliographystyle{plain} 

\bibliography{citations}      




\end{document}